# Ultrabroadband Integrated Photonics Empowering Full-Spectrum Adaptive Wireless Communications

Zihan Tao[1,†], Haoyu Wang[1,†], Hanke Feng[2,†], Yijun Guo[1,†], Bitao Shen[1,†], Dan Sun[3], Yuansheng Tao[2], Changhao Han[4], Yandong He[5], John E. Bowers[4], Haowen Shu[1,6,*], Cheng Wang[2,*] & Xingjun Wang[1,3,6,*]

[1]State Key Laboratory of Advanced Optical Communications System and Networks, School of Electronics, Peking University, Beijing, China.
[2]Department of Electrical Engineering State & Key Laboratory of Terahertz and Millimeter Waves, City University of Hong Kong, Kowloon, China.
[3]Peking University Yangtze Delta Institute of Optoelectronics, Nantong, China.
[4]Department of Electrical and Computer Engineering, University of California Santa Barbara, Santa Barbara, CA, USA.
[5]School of Integrated Circuits, Peking University, Beijing, China.
[6]Frontiers Science Center for Nano-optoelectronics, Peking University, Beijing, China.
[†]These authors contributed equally to this work.

Corresponding authors: *haowenshu@pku.edu.cn, *cwang257@cityu.edu.hk, *xjwang@pku.edu.cn

**Abstract**

**The forthcoming sixth-generation (6G) and beyond (XG) wireless networks are poised to operate across an expansive frequency range–from microwave, millimeter-wave to terahertz bands–to support ubiquitous connectivity in diverse application scenarios. This necessitates a one-size-fits-all hardware solution that can be adaptively reconfigured within this wide spectrum to support full-band coverage and dynamic spectrum management. However, existing electrical or photonic-assisted wireless communication solutions see significant challenges in meeting this demand due to the limited bandwidths of individual devices and the intrinsically rigid nature of their system architectures. Here, we demonstrate adaptive wireless communications over an unprecedented frequency range spanning over 100 GHz, driven by a universal thin-film lithium niobate (TFLN) photonic wireless engine. Leveraging the strong Pockels effect and excellent scalability of the TFLN platform, we achieve monolithic integration of essential functional elements, including baseband modulation, broadband wireless-photonic conversion, and reconfigurable carrier/local signal generation. Powered by broadband tunable optoelectronic oscillators, our signal sources operate across a record-wide frequency range from 0.5 GHz to 115 GHz with high frequency stability and consistent coherence. Based on the broadband and reconfigurable integrated photonic solution, we realize, for the first time, full-link wireless communication across 9 consecutive bands, achieving record lane speeds of up to 100 Gbps. The real-time reconfigurability further enables adaptive frequency allocation, a crucial capability to ensure enhanced reliability in complex spectrum environments. Our proposed system marks a significant step towards future full-spectrum and omni-scenario wireless networks.**

## Introduction

Wireless communication technology has profoundly influenced our information society by enabling widespread and massive connectivity[1,2]. To meet the growing demand for ubiquitous access, future sixth-generation (6G) and beyond (XG) networks are anticipated to adaptively utilize full-spectrum resources for diverse application scenarios[3–5], as summarized in Fig.1a. For example, high-frequency millimeter-wave and sub-terahertz bands will provide further increased data speed and reduced latency, facilitating emerging data-intensive services like extended reality (XR) and remote surgery[6]. Meanwhile, the low-loss sub-6 GHz and microwave bands continue to provide wide spatial coverage in rural areas or urban centers[5,7–9]. Furthermore, the system should feature real-time spectral reconfigurability to ensure efficient spectrum usage and reliable access in complex spectrum environments, commonly referred to as 'intelligent radio'[10–12]. To support this adaptive full-spectrum vision, a one-size-fits-all hardware solution that can be reconfigured to operate within the entire spectrum is much desired[13]. Specifically, it should support high-fidelity and broadband conversion between baseband and radio frequency (RF) bands, low-noise signal sources with wideband tunability and consistent performance, as well as low-cost chip-scale integrability that seamlessly combines all these essential functions in a small form factor for synergistic operation.

However, achieving such a broadband and reconfigurable hardware presents significant challenges[14]. Traditional electrical solutions are typically designed for operation in a particular band only[15], as the underlying electrical components require distinct design rules, structures, and materials for each band. Consequently, covering a broad frequency spectrum requires a series of independent subsystems specially tailored for respective bands. This not only increases system complexity and cost especially in large-scale deployment scenarios[16,17], but also limits the frequency reconfigurability needed for dynamic spectrum management. Moreover, electrical signal sources based on cascaded frequency multipliers see significantly increased noise at high frequencies, leading to inconsistent performances among different bands and degraded communication quality[18,19]. Recent advances in optoelectronic technology present new opportunities to achieve broadband operation by leveraging the inherently large bandwidth of optical systems[20–29]. Specifically, multi-band wireless reception can be achieved optically through broadband electro-optic (E-O) modulators, i.e. wireless-photonic conversion[28]. The modulated optical signals can be further flexibly processed using optical apparatus and distributed over fiber networks[30]. On the signal generation front, wireless carrier and local oscillation signals can be generated via optical downmixing of two laser sources, where the RF signal frequency can be flexibly set by controlling the laser frequency and the tuning range is only bounded by the photodetector bandwidth[29]. This approach further allows high-speed in-phase/quadrature (I/Q) modulation of the wireless signal using mature optical modulation techniques, supporting ultrahigh data throughput at low cost.

Despite being naturally high-frequency friendly, achieving a compact and multi-band adaptive photonic wireless system faces three key challenges. First of all, current optical wireless generation schemes struggle to achieve low noise, wideband tunability, and system compactness simultaneously. Signals generated from beating two unrelated free-running lasers generally exhibit large phase noises and frequency instability. Better coherence can be achieved by beating two optical sidebands that originate from a single laser source, when modulated by an external microwave source[21,31]; however, such optical frequency multiplication process is still constrained by the same noise accumulation law as that in the electronic multiplier case. While more advanced approaches like optical frequency division (OFD) can offer extremely low phase noise[25–27], they often require complex external locking schemes with compromised spectral flexibility and system simplicity. Secondly, wireless-photonic conversion in current demonstrations is often realized using bulk lithium niobate[31] or silicon-based modulators[28,32], which suffer from high drive voltages, or large intrinsic optical insertion loss caused by their intrinsic mechanisms such as plasma dispersion or surface plasmon polariton effect. These limitations degrade the signal-to-noise ratio (SNR) and signal speed. Finally, although some of these demonstrations partly incorporate integrated photonic technologies, the full communication systems remain largely discrete. In short, the quest for a low-noise, wideband tunable, and highly integrated wireless solution based on photonic integrated circuits (PIC) remains elusive.

Here, we address these challenges by developing a universal photonic wireless engine with broadband reconfigurability based on a thin-film lithium niobate (TFLN) platform[33–39]. Compared with traditional LN devices, TFLN platform offers E-O modulators with much broader bandwidths covering millimeter-wave and sub-THz bands, as well as substantially improved scalability to monolithically integrate multiple E-O and linear functional devices on a single chip[33,39]. Leveraging these unique features, we achieve ultrabroadband wireless carrier/local oscillation (LO) signal generation based on an optoelectronic oscillator scheme[19,40,41], which produces highly stable microwave signals with a record-broad frequency tuning range from 0.5 GHz to 115 GHz while maintaining consistent coherence. We further integrate the broadband signal sources with baseband modulation and wireless-photonic conversion modules to realize a fully functional integrated wireless-photonic circuit that enables high-fidelity adaptive wireless communication across a bandwidth exceeding 100 GHz. We achieve, for the first time, end-to-end high-speed wireless communication across 9 consecutive bands, with a peak data rate exceeding 100 Gbps. Based on the ultrabroad operation bandwidth and real-time reconfigurability, we showcase system-coordinated spectrum management that successfully addresses three classic wireless channel quality improvement challenges— namely, channel self-adaptation, interference avoidance, and dynamic homodyne alignment. Our ultra-broadband wireless-photonic solution could offer unique reconfigurability and adaptivity for next-generation 'intelligent radio' wireless networks.

**Results**

**Universal PIC-based wireless system architecture**

Fig.1 b,c illustrate the schematics of our proposed broadband reconfigurable wireless-photonic solution, where key functional elements, including carrier/LO signal generation, wireless-photonic conversion, and baseband modulation, are integrated on the same TFLN chip for both wireless signal transmission and reception. The photonic wireless engine is monolithically fabricated based on a wafer-scale stepper lithography process (see supplementary note I), and packaged for functional system characterizations.

At the wireless transmitter end (Tx), a broadband tunable carrier is generated using an optoelectronic oscillator comprising a high-speed E-O phase modulator and a high quality (Q) microring resonator (MRR). At the intended optoelectronic oscillation frequency, one of the two phase modulated sidebands is aligned with the resonance frequency of the MRR and filtered out, leading to phase-to-intensity modulation conversion. The modulated optical signal is converted back to the electrical domain at the photodetector, which is finally sent back to the E-O modulator to close the optoelectronic oscillation loop. With sufficiently high gain from the optical and electrical amplifiers, positive feedback can be achieved leading to self-oscillation within the optoelectronic loop. Leveraging an advanced slotted electrode design, our E-O modulator exhibits only 1.3 dB E-O roll-off at 67 GHz, with an extrapolated 3-dB bandwidth of 110 GHz. This allows effective excitation of optoelectronic oscillation within an unprecedented bandwidth. The MRR features a high intrinsic Q factor of nearly 1 million and a free spectral range (FSR) of 134 GHz, enabling narrowband optical filtering within a wide side-mode-free frequency range. On-chip thermal tuning of this MRR filter leads to reconfigurable selection of the oscillating frequency. To load the baseband signal onto the wireless carrier, an in-phase quadrature (IQ) modulator is first deployed to convert the signal to the optical domain via carrier-suppressed single-sideband modulation. Afterwards, the IQ-modulated sideband is mixed with the filtered optoelectronic oscillation signal at a broad-bandwidth photodetector. This generates a modulated wireless signal with a carrier frequency determined by the optoelectronic oscillation frequency. Importantly, the carrier generation and IQ modulation processes share the same laser source, which ensures strong coherence in the final generated signal. Altogether, the Tx chip generates an uplink wireless data stream with a tunable center frequency across a broad bandwidth.

At the receiver end (Rx), an optically assisted wireless signal down-conversion is performed, which eliminates the need for band-specific RF mixers. To achieve this, we use another wideband E-O modulator with a low half-wave voltage ($V_\pi$) of 2.8 V to directly convert wireless signals from the receiving antenna to the optical domain. To convert the modulated optical signal back to baseband, a tunable optical LO signal is generated based on the same optoelectronic oscillation process as that used in the Tx. The modulated sideband is filtered and finally mixed with the optical LO at a receiving photodetector to retrieve the baseband communication signal. The mirrored system architecture for carrier and LO generation ensures broad bandwidths and reconfigurability at both Tx and Rx ends. Moreover, the LO frequency can be fine-tuned to precisely align with the center frequency of the received signal, enabling zero intermediate frequency (IF) reception. The zero-IF signal and LO can be either directly demodulated on-site, or further distributed to a remote central unit (CU) through low-loss fiber networks. Apart from broad bandwidth and low $V_\pi$, the TFLN E-O modulator also features a high modulation linearity, benefiting from the linear Pockels effect of lithium niobate, which ensures minimal signal distortion during the E-O conversion process. The measured spurious-free dynamic range (SFDR) of the modulator is approximately 99 dB·Hz$^{2/3}$, which is limited by the sinusoidal transfer function of the Mach-Zehnder interferometer and can be further improved using advanced linearization strategies[42]. The experimental details and performance characterization results are provided in Supplementary Note II.

Overall, the proposed photonic wireless core minimizes the use of bandwidth-limited devices, requiring only essential electrical amplifiers and wireless antennas as peripheral circuits. By working with various specific peripheral circuits, it ultimately meets the demands of 6G omni-scenarios, supporting full-band coverage and adaptive wireless communication.

**Consistent operation across ultra-broad bandwidth**

We first show that individual elements of the proposed integrated photonic wireless engine provide high and consistent performance across a broad bandwidth. For wireless-photonic conversion (Fig.2a), wireless waveforms with different carrier frequencies under quadrature phase shift keying (QPSK) modulation are transmitted and received by the antenna-linked modulator. The measured optical spectra are shown in Fig.2b, revealing minimal intensity variations in the received optical sidebands at modulation frequencies up to 100 GHz. Further characterization of the high frequency (80 GHz) wireless receiving response within a wide optical carrier wavelength range from 1515 nm to 1630 nm (Fig.2c) also demonstrates uniform optical spectral performance. The consistency across both wireless carrier and optical frequency bands highlights its potential for wavelength division multiplexing-based high capacity radio-over-fiber fronthaul system[30].

We next demonstrate carrier and LO signal generation with flexible reconfigurability and low-noise performance across a record-wide frequency range from 0.5 to 115 GHz, using the setup illustrated in Fig.2d (See Methods for details). The optoelectronic oscillation process follows the Van der Pol model[43], where the oscillation frequency is determined by the relative frequency difference between the laser and the resonance peak of MRR. Fig.2g presents the measured signal spectra, with nearly continuous coverage of the full frequency range with a 500 MHz frequency interval. The two narrow spectral regions with missing data correspond to oscillation frequencies close to half the MRR's FSR (∼ 67 GHz) and in between two types of amplifiers (∼ 76 GHz), respectively. This E-O hybrid signal generator effectively unites 9 RF frequency bands across more than 7 octaves in frequency, based on a universal photonic chip solution, a feat untenable for traditional electronics. Although dedicated peripheral elements like electrical amplifiers are still required for respective bands, these relatively narrowband elements naturally mitigate the impact of adjacent resonances from the optical MRR (as illustrated in Fig.2e). As a result, this alleviates the demand for extremely small MRRs to increase the frequency separation between resonant peaks[41].

More importantly, the proposed signal generation architecture overcomes the critical noise-accumulation challenge in conventional frequency multiplier schemes (Fig.2f), offering minimized and consistent noise at high frequencies. Fig.2h displays the measured phase noise curves at different generated frequencies of 5 GHz, 20 GHz and 90 GHz, which align closely with each other with consistent phase noises of -85 dBc/Hz at 10 kHz frequency offset (green curves). This can be further reduced by incorporating a longer fiber loop. The measured phase noise at 10 kHz frequency offset with a 2 km fiber delay is substantially reduced to around -110 dBc/Hz at 10 kHz (blue curves). Here the side modes are not effectively suppressed due to the relatively wide filtering bandwidth of the MRR, which can be addressed by employing the dual-loop method[44,45] and further improving the Q factor of MRR[19]. The measured phase noise at 90 GHz rises beyond 10 kHz offset due to additional noise from the down-mixing detection system used (see supplementary note III). Further, to characterize the stability of the platform and structure, the frequency drift of the free-running oscillation at around 20 GHz was recorded every minute for an hour by an electrical spectrum analyzer. As shown in Fig.2i, the results indicate that majority of the deviations are less than 0.5 parts per million (ppm), which is nearly 10 times better than other platforms even under PID-based MRR's frequency locking[46].

**Multi-band converged wireless communications**

We next leverage the broadband consistency of the proposed photonic wireless system to demonstrate end-to-end wireless communication reconfigurable across 9 consecutive frequency bands. Fig.3a shows a detailed system configuration of the Tx and Rx ends, where the broadband reconfigurable photonic wireless core co-operates with supporting electronics and antennas in respective bands to achieve full system functions (See Methods for detailed description). Compared with previous photonic-assisted wireless communication methods based on silicon-based modulators, the TFLN modulators here provide lower insertion loss (less than 2 dB), lower half-wave voltage, and larger bandwidths, therefore ensuring minimal power penalty and good SNR across an ultrabroad spectrum range (see supplementary note IV for detailed analysis). To validate these advantages, we conduct full-spectrum wireless communications at carrier frequencies from 5 GHz to 100 GHz with a 5 GHz interval, utilizing horn antennas for the transmission and reception of high-speed wireless signals. Three sets of antennas are used to cover specific bands: 2-18 GHz, 18-50 GHz, and 75-110 GHz. The absence of the 50-75 GHz band is due to its limited use, as atmospheric absorption is substantial within this range. Fig.3b displays representative constellation diagrams in different frequency channels, with various data rates and modulation formats, i.e. QPSK and 16 quadrature amplitude modulation (16-QAM). Up to 100 Gbps single-lane transmission is achieved at both 35 GHz and 95 GHz center frequencies (with even lower bit error ratio (BER) at 95 GHz), which presents the highest data rate for integrated photonics-assisted wireless communication. Moreover, apart from the lower-frequency carriers with limited bandwidths, all channels with carriers above 30 GHz achieved data rates exceeding 50 Gbps, validating the system's performance consistency across a wide frequency range.

We summarize the BER results in all measured bands in Fig.3c, which are obtained using a standard digital signal processing flow (see Methods for more details). All transmission BER values, across 9 different bands (including L, S, C, X, Ku, K, Ka, U, and W), remain below the hard-decision forward error correction (HD-FEC) or soft-decision forward error correction (SD-FEC) thresholds. Currently, the data transmission performance is mainly limited by two factors. Firstly, the performances of the antennas and amplifiers are optimized at their respective center frequencies and degrade away from them. This can be clearly seen in the BER values of 60 Gbps transmission results in the W band (Red dashed line in Fig.3c), which substantially deteriorates near the band edges. The second limiting factor is the non-ideal in-band spectral response of electrical devices, i.e. ripples and dips, which induces uneven sideband loading. This type of signal degradation is more pronounced for higher-order modulation formats, such as 16-QAM[47], and can be mitigated through cooperative fine-tuning of the carrier and LO frequency, as discussed in the next section.

**Dynamic spectrum management**

We finally demonstrate the system's real-time wideband reconfigurability for dynamic spectrum management. Leveraging the thermo-optic effect, the system is capable of rapid carrier/LO frequency adjustments, achieving a 6 GHz tuning range within 180 µs (see Supplementary Note V for details). This capability greatly enhances the adaptability of wireless systems in complex real-world scenarios. For example, as illustrated in the previous section, the non-ideal responses of electrical devices, primarily the antenna and electrical amplifier, create significant fluctuations within certain bands. Additional effects such as multipath interference[48], can also introduce similar deterioration, as described in Fig.4a. Traditionally this effect is usually mitigated by orthogonal frequency-division multiplexing (OFDM) algorithms[49]. Here we provide a more fundamental hardware solution, by adaptively search for an optimal frequency point with enhanced reliability through near-continuous frequency-domain tuning. As shown in Fig.4b, the LNAs used at both Tx and Rx ends in the W band exhibit uneven frequency responses with amplitude variations exceeding 4 dB (see supplementary note VI for detailed information). Operating at 80 GHz, this fluctuation induces noticeable distortions in the received optical spectrum (see Fig.4c.i). Adaptively shifting the operating frequency to 98 GHz significantly improves signal quality, yielding a flatter spectral envelope in the optical domain (Fig.4c.ii). Fig.4d summarizes the measured BER for 120 Gbps 16-QAM signals between 94 and 100 GHz, with an optimized operation point at 97.5 GHz, confirming the effectiveness of reconfigurable systems in achieving channel self-adaptation.

Another common challenge in wireless communication is interference due to noisy and congested wireless environment. As shown in Fig.4e, an additional antenna is introduced to emit an interference signal that overlaps with the original signal in the frequency domain. Both signals are received simultaneously by the Rx antenna. Based on the reconfigurable photonic architecture, the transmitted signal at the Tx end can be adaptively tuned to avoid the jammed band and make use of a free frequency band (See supplementary note VII for detailed analysis). Importantly, the LO at the Rx end can be synchronously reconfigured to maintain frequency alignment during this process (to be discussed next). In the experiment, a 20 Gbps QPSK signal serves as the target signal, while a microwave source (Keysight 8257D) generates single-tone signals with different amplitudes and frequencies as interference. Adaptable communication at 25 GHz and 96.55 GHz is demonstrated as shown in Fig.4f and g, respectively. In both cases, the interference signal initially overlaps with the data signal, which completely disrupts the received signals with unrecoverable constellation diagrams. Properly tuning the center frequencies of Tx and Rx in both cases successfully suppressed the interference with a baseband low-pass filter, resulting in a BER of 0. The end-to-end reconfigurability of the system significantly improves communication reliability in complex electromagnetic environments.

Finally, we emphasize the importance of system-coordinated reconfigurability at both Tx and Rx ends. To address the above challenges based on adaptive spectrum management, a crucial requirement is that the carrier frequency at Tx and the LO frequency at Rx must be precisely aligned to achieve zero-IF (Fig.4h), as any frequency offset can adversely deteriorate the communication quality. Fig.4i illustrates the measured BER values for various signal speeds and modulation formats when LO is detuned from the carrier frequency. Compared with QPSK, 16-QAM is more sensitive to the frequency offset, highlighting the need for precise carrier-LO matching for more advanced modulation formats. When the frequency offset becomes too large for the DSP to accurately estimate, the signal completely fails to recover, resulting in a BER of 0.5.

**Conclusion and discussion**

**Table.1** Comparison of representative works in photonic-assisted high-speed wireless communication

| Year | Key device | Developed Functionalities | Operational Range (GHz) | Dynamic Response Speed (Frequency/Time) | Verified Channels Count | Single-channel speed (Gbps) | Dynamic Spectrum Management |
|---|---|---|---|---|---|---|---|
| 2013[23] | UTC[1]-PD | Mixer | Fix | N/A[4] | 1 | 100 | × |
| 2019[28] | Plasmonic-Mod[2] | Wireless-Photonic Conversion | Fix | N/A | 1 | 50 | × |
| 2022[20] | OFC[3] | Carrier Generation | Fix | N/A | 1 | 131 | × |
| 2023[24] | OFC | Carrier Generation | Fix | N/A | 1 | 60 | × |
| 2024[21] | Bulky LN-Mod | Wireless-Photonic Conversion | Fix | N/A | 1 | 32 | × |
| This Work | Integrated Circuit in TFLN | Full link | 0.5-115 | 6 GHz/180 µs | 14 | 100 | ✓ |

[1] UTC:uni-travelling-carrier. [2] Mod:Modulator. [3] Opitcal frequency comb. [4] N/A:not applicable.

In summary, we propose and demonstrate an integrated optoelectronic architecture to unite wireless spectrum from 0.5 to 115 GHz for multi-band converged wireless communications. Fundamental components such as carrier and LO generation, signal loading and receiving are implemented on the same TFLN platform that features broadband photonic building blocks and excellent scalability. This enables reconfigurable full-link wireless communications with unprecedented bandwidth, data rates, and system functionality compared with previous photonic-assisted wireless demonstrations, as summarized in Table.1. The wideband consistency also enables the proposed system to adapt effectively to complex electromagnetic environments, further enhancing its reliability for real-world wireless communication.

Currently, the wireless communication system is demonstrated in an indoor environment with a wireless propagation distance of 1.3 m, limited by the low-gain antenna (below 20 dB in the W band) and experimental conditions. Long-distance wireless propagation can be achieved with appropriate antennas and focusing lenses. The integration level and performance can be further enhanced through advanced integration techniques[50]. The operational bandwidth could be extended to the terahertz (THz) range using ultra-broadband TFLN modulators and modified uni-traveling-carrier (MUTC) photodetectors[51,52]. Recent advancements in heterogeneously integrated III-V-on-TFLN technologies could allow high-coherence lasers via self-injection locking[53] and photodetectors[54] integrated on the same TFLN chip. Furthermore, the Q factors of the MRRs in TFLN can be further enhanced[55], enabling higher resolution in spectral filtering.

Moving forward, the proposed system has the potential to serve as a universal photonic wireless engine, where AI algorithms can be implemented to dynamically adapt the hardware to changing environments and network dynamics, following the "AI native" concept[11]. Moreover, the proposed schematic can be seamlessly adapted for integrated sensing and communication (ISAC)[56], where the loaded signal can further integrate linear frequency modulation (LFM) signal, towards real-time data transmission and precise environmental sensing at the same time.

## Methods

### The principle and experimental details of OEO

The proposed architecture in the experiment employed the phase modulation to intensity modulation (PM-IM) principle, utilizing components including an on-chip phase modulator, MRR, EDFA (Amonics, AEDFA-PA-35-B-FA), PD, and electrical amplifier. Throughout the experiment, we used a thermoelectric cooler (TEC, Vescent) to regulate the chip temperature. For the 0.5-50 GHz range, we use the low-noise amplifiers (from low frequency to 50 GHz) with an RF power divider to monitor the oscillation frequency. In the 50–70 GHz range, we replaced the corresponding peripheral components with WR-15 waveguide-interface low-noise amplifiers and a WR-15 waveguide-interface RF power divider, supporting frequencies from 50 to 75 GHz. The frequency void in the range of 64-67 GHz is equivalent to half of the FSR of the MRR. This phenomenon occurs due to the gain competition between two adjacent resonance peaks. This can be solved by coordinating the MRR's FSR with the amplifier's bandwidth. For the high-frequency range (75 GHz and beyond), we substitute the peripheral components with WR-10 waveguide-interface amplifiers and a WR-10 waveguide-interface RF power divider. The inability to oscillate at 74-76 GHz is due to low gain at the edge of the electrical amplifiers' bandwidth, which can be resolved by using an amplifier capable of covering this frequency range. We collect data for 0.5-50 GHz using an electrical spectrum analyzer (Keysight N9021B). For 50-90 GHz, we use frequency extension modules (Keysight M1971V, Keysight M1971E Opt003) to extend the measurement frequency range. To obtain data for the 90–115 GHz range, we mix the generated signal with an 80 GHz single tone generated by a microwave source (Keysight E8257D) with a x4 frequency multiplier. In Fig.2g, each color-coded data collection covers a 10 GHz range. The collection range for the 0-10 GHz band is limited to 1 GHz to prevent the second harmonic from affecting visualization. The difference in noise floor levels across frequencies is mainly due to the varying response of the spectrum analyzer across different frequency bands. More details about the experimental setup can be found in Supplementary note III.

### Details of Experimental Communication Link

The full spectrum wireless communication is conducted using the independent transmitter (Tx) and receiver (Rx) end. Three sets of antennas are used to cover specific bands: 2-18 GHz, 18-50 GHz, and 75-110 GHz, while the amplifiers used in the wireless communication link cover the ranges of 0-40 GHz and 40-67 GHz, respectively. In the Tx end, the in-phase quadrature (IQ) modulator and optoelectronic oscillator (OEO) without additional fiber delay line are driven by the same local laser ($f_o$) with a 50% power splitter to implement baseband modulation and carrier generation, respectively. Specifically, to eliminate the impact of bias point instability on communication quality, we utilized a commercial LN-based IQ modulator equipped with a built-in bias controller to ensure that the modulator operates in a carrier suppression single-sideband modulation state. Then the baseband modulation is implemented by injecting I, Q signal generated through AWG. The TFLN chip used for OEO integrates a phase modulator and a microring resonator (MRR) for electro-optic conversion and mode selection. The light coupled out from the chip is amplified by an erbium-doped fiber amplifier (EDFA) and split by a 50% power splitter. One path is directed to a photodetector (PD) for conversion back to an electrical signal, which is amplified and fed back to the on-chip modulator for single-mode oscillation with center frequency $f_c$. The other path passes through an optical bandpass filter (BPF) to select the sideband that has not been filtered by on-chip MRR, which has a center frequency of $f_o + f_c$. The sideband extracted by BPF is then combined with the output of the IQ modulator and photomixed in PD to convert into electrical waveforms, whose center frequency is represented as $f_e = f_c$ and spectrum width is the same as the IQ signal generated through AWG. The waveform is electrically amplified and emitted by the antenna. After 1.3 m of wireless propagation, the waveform is received in the Rx end by the second antenna.

In the Rx section, the Rx modulator and the second OEO are driven by the same laser ($f_{o2}$). We use a phase modulator for wireless-photonic conversion as it still has good linearity without bias control, which simplifies the complexity of the end-to-end experiment. The signal coupled out from the phase modulator is filtered to reserve one single sideband and amplified by an EDFA. The signal coupled out from the phase modulator is processed through a band-notch filter (BNF) to suppress the carrier. Subsequently, the filtered signal is amplified using EDFA before being fed into the coherent receiver. The Rx end OEO has the same configuration as the Tx end OEO, and is set to oscillate at the same frequency $f_c$ as the Tx end OEO. The light coupled out from Rx OEO passes through another BPF to select the sideband that has not been filtered by on-chip MRR. After amplification, this sideband can serve as the local oscillator for the demodulation of the signal. Along with the sideband-generated phase modulator, it is sent to a coherent receiver for demodulation and signal recovery.

**Digital signal processing**

At the Tx, random binary bit streams are mapped into QAM symbols. The QAM symbols are up-sampled and subjected to root raised cosine pulse shaping with a roll-off factor of 0.05. The in-phase and quadrature paths of the QAM signal are loaded into the AWG and then transmitted. At the receiver side, Gram-Schmidt orthogonal normalization is first performed, followed by matched filtering. Subsequently, the signal is down-sampled from 128 GSa/s to two samples per symbol. Adaptive equalization and carrier recovery are then performed. The equalizer has a 2x4 MIMO structure. The equalizer's tap coefficients are pre-converged using the constant modulus algorithm, and the pre-converged output is used for frequency offset estimation. Frequency offset estimation is achieved by taking the fourth power of the signal and identifying the highest spectral peak. Afterward, the equalizer in DD-LMS mode is employed, and carrier phase estimation and compensation are performed within the equalizer's update loop using the blind phase search algorithm with 16 test angles. After performing equalization and carrier recovery, orthogonalization is applied to mitigate hybrid imperfections and modulator bias problems. Subsequently, symbol decisions are determined. BERs and EVMs are calculated to evaluate the system's performance.

**Data availability**

The data that supports the plots within this paper and other findings of this study are available from the corresponding authors upon reasonable request.

**Code availability**

The codes that support the findings of this study are available from the corresponding authors upon reasonable request.

**Acknowledgments**
This work was supported by National Key Research and Development Program of China (No. 2022YFB2803700), National Natural Science Foundation of China under Grant (62235002, 62327811, 62322501, 623B2004, 12204021), the Research Grants Council, University Grants Committee (CityU 11212721, CityU 11204022, C1002-22Y), and Croucher Foundation (9509005). The authors thank the technical support of Mr. Chun Fai Yeung, Mr. C.W Lai, Miss. Olive Chan, and Mr. Li Ho at HKUST, Nanosystem Fabrication Facility (NFF) for the stepper lithography and PECVD process. The authors also thank Miss. Shiwen Xu for manuscript polishment.

**Author contributions**

The experiments were conceived by Z.T., H.W., H.F. and Y.G. The TFLN-base devices were designed by Z.T. and H.F. The chip was fabricated by H.F. The experiments were performed by Z.T., H.W., Y.G. and H.F. The results were analyzed by Z.T., H.W., Y.G. and H.F, with the assistant from B.S., Y.T., and C.H. The chip-scale co-package was performed by D.S. The DSP algorithm was implemented by Y.G. and H.W. All authors participated in the writing of the manuscript. The project was under the supervision of J.E.B, H.S, C.W. and X.W.

**Competing financial interests**

H.F. and C.W. are involved in developing lithium niobate technologies at Kokoxili Photonics Limited.

**Additional information**

Supplementary information is available in the online version of the paper. Reprints and permissions information is available online. Correspondence and requests for materials should be addressed to Haowen Shu, Cheng Wang, and Xingjun Wang.

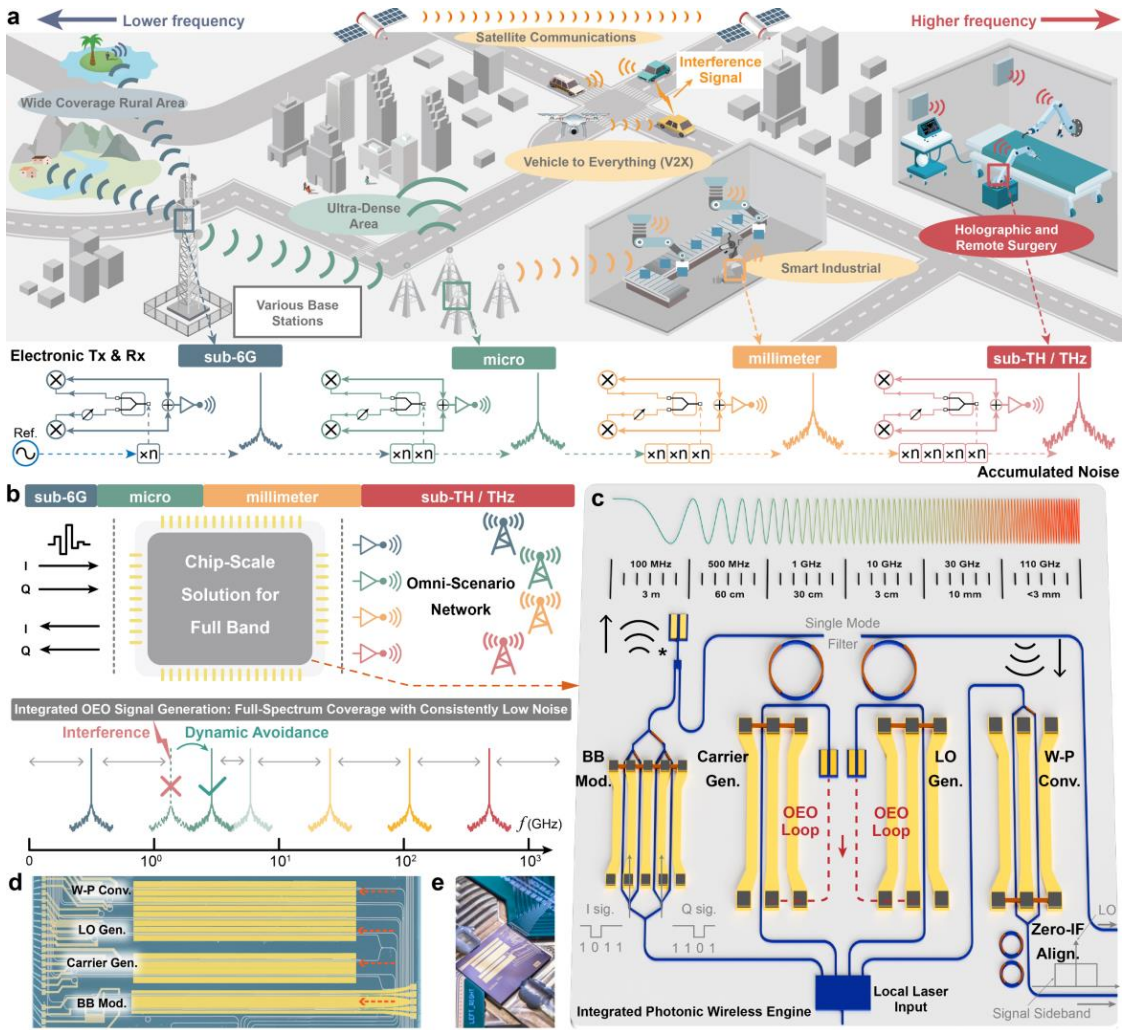

**Figure 1 Ultrabroadband integrated photonics empowering omni-scenario wireless networks. a,** Envisioned omni-scenario wireless networks with a large frequency span from sub-6 GHz low frequency (dark green) to sub-THz high-frequency (red) bands. Traditional electrical schemes require distinct device sets for each band, with accumulated noises from the multiplier-based sources at high frequencies. **b,** Conceptual diagram of the proposed integrated photonic scheme enabling adaptive full-spectrum operations. A broadband optoelectronic oscillator generates frequency-tunable signals covering the full band with a consistently low noise level. **c,** Schematic of the thin-film lithium niobate photonic wireless solution for ultra-broadband carrier and local oscillator generation, signal modulation, and reception. * Components currently not integrated on the TFLN-chip. **d,** Optical microscope image of the fabricated TFLN chip. **e,** Photo of the co-packaged photonic-wireless system. BB Mod.: baseband modulator; W-P Conv.: wireless-photonic conversion; LO: local oscillation; Gen.: generation; OEO: optoelectronic oscillation. Zero-IF Align.: zero-intermediate frequency alignment.

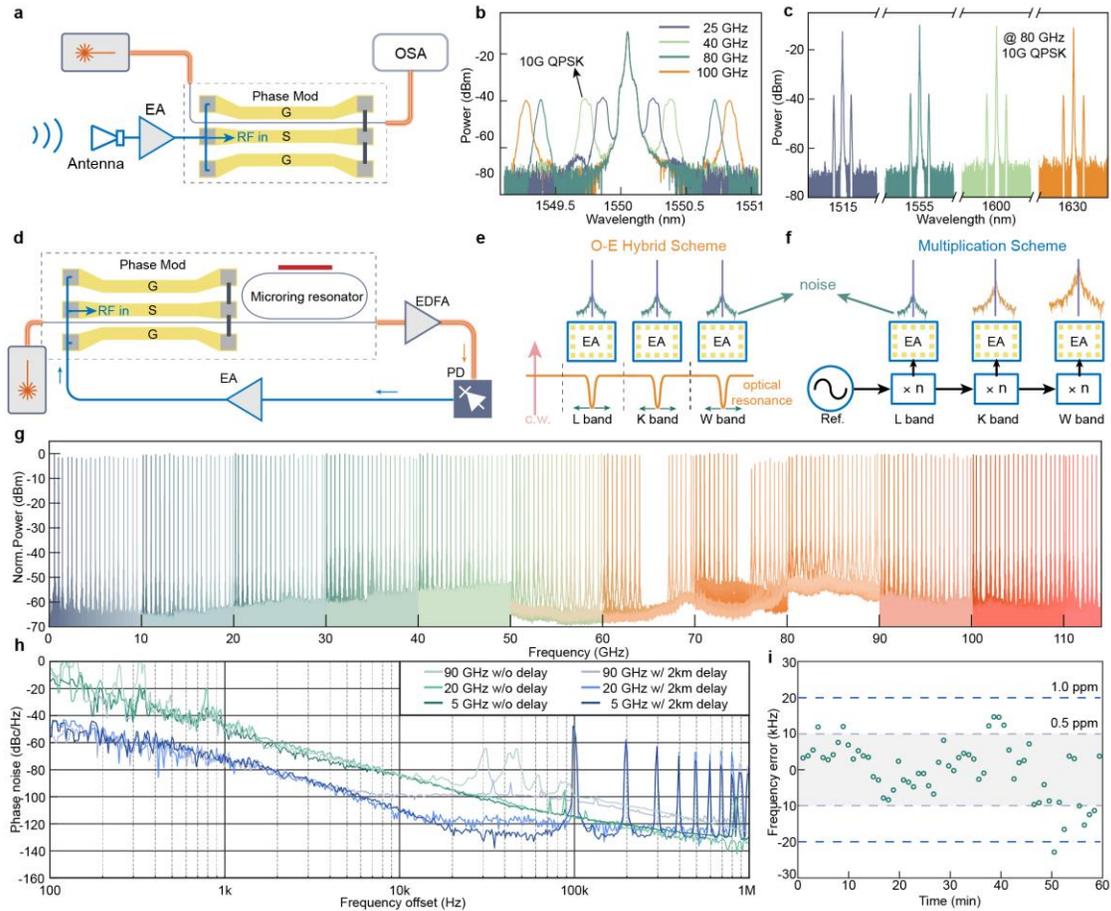

**Figure 2 Broadband wireless-photonic conversion and wireless signal generation. a,** Setup for wireless-photonic conversion. **b,c,** Measured optical spectra at various wireless (b) and optical (c) carrier frequencies. **d,** Setup for carrier signal generation based on optoelectronic oscillation (OEO). **e,** Working principle of the OEO scheme, where the oscillation frequency can be reconfigured by tuning the optical resonance frequency, without affecting noise performance. **f,** Working principle of traditional multiplier-based signal sources, where the noise cascades. **g,** Measured OEO spectrum with a resolution bandwidth (RBW) of 51 kHz. **h,** Measured phase noises of OEO signals at various frequencies showing consistent performances across bands. **i,** Measured frequency error diagram. OSA: optical spectrum analyzer; EDFA: erbium-doped fiber amplifier; EA: electrical amplifier; PD: photodetector; MS: microwave source; c.w.: continuous wave.

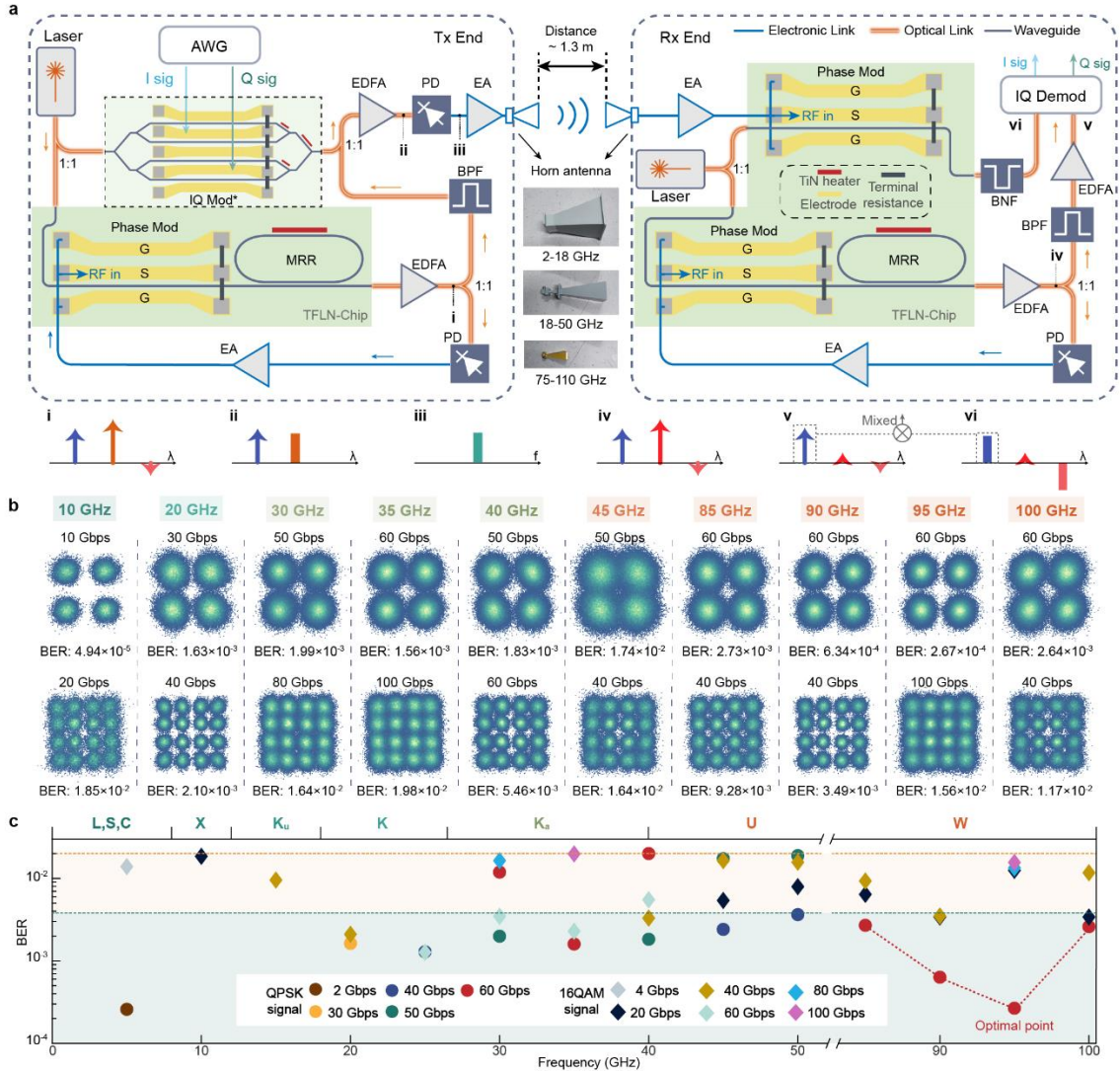

**Figure 3 Multi-band wireless communication results. a,** Setup for end-to-end wireless communication experiments. Inserts (i-vi) show the optical and electrical spectra at critical locations of the circuit. **b,** Measured constellation diagrams in different frequency bands. **c,** Summary of the measured bit-error ratios (BERs) in different frequency bands. Orange and blue dashed lines correspond to the hard-decision and soft-decision forward error correction thresholds, respectively. AWG: arbitrary waveform generation; MRR: microring resonator; IQ Mod*: In-phase and Quadrature Modulator. We utilized a commercial IQ module with integrated bias point control to currently replace the on-chip IQ modulator, which simplified the complexity of the communication experiment and prevented additional imbalance; EDFA: erbium-doped fiber amplifier; EA: electrical amplifier; BPF: optical band-pass filter; BNF: optical band-notch filter.

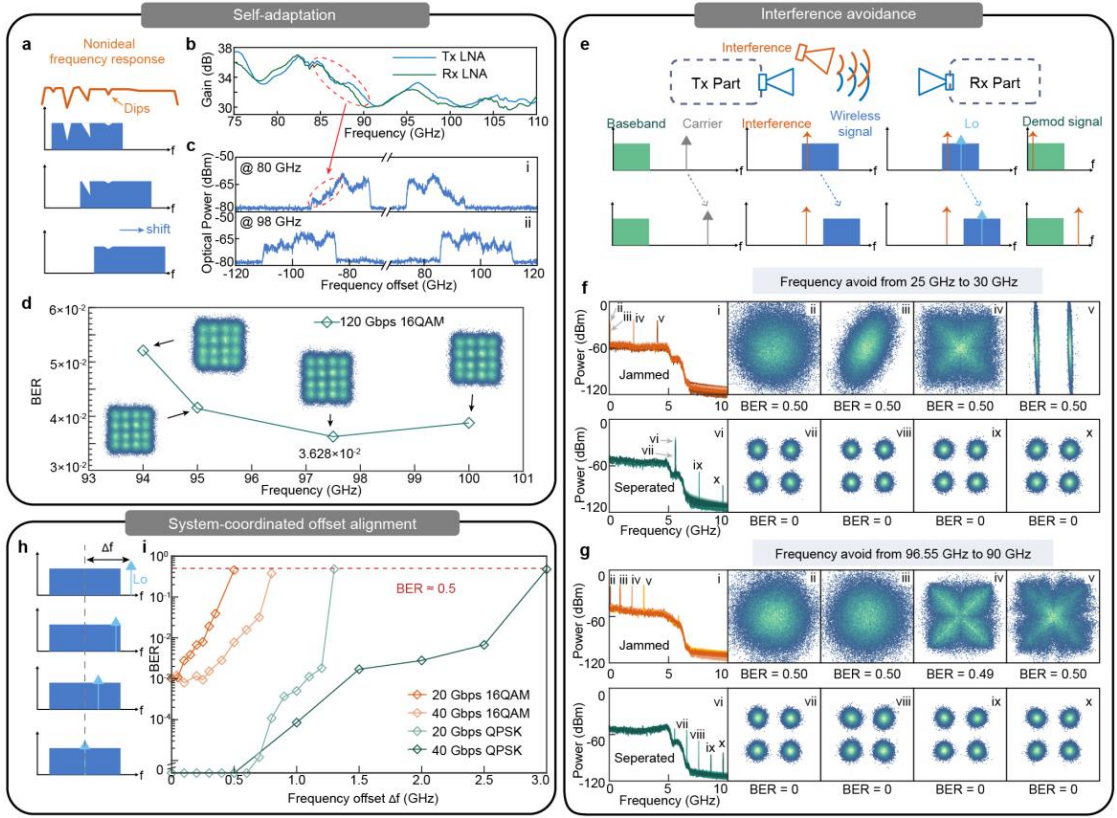

**Figure 4 System-coordinated wireless spectrum management. a,** Schematic diagram of spectral self-adaptation. **b,** Non-ideal frequency response of the wireless channels in the experiment. **c,** Optical spectra distortion of the loaded signals for wireless carrier frequencies of 80 GHz (i) and 98 GHz (ii), respectively. **d,** Measured BERs of 120 Gbps 16-QAM signal in different frequency channels. **e,** Schematic illustrations of the setup and principle of active interference avoidance. **f,g,** Baseband power spectra and demodulated constellation diagrams in the cases of jammed (insets i-v) and separated (insets vi-x) signals at initial frequencies of 25 GHz (f) and 96.55 GHz (g), respectively. Insets (i) and (vi) show the power spectral density of the target and interference signals in respective cases. Insets (ii-v,vii-x) show demodulated constellation diagrams and BERs. h, Schematic diagram of frequency alignment between target signal and LO. i, BER versus frequency offset for four types of signals.